\documentclass[aps,pre,twocolumn,groupedaddress]{revtex4-2}

\usepackage{color}

\usepackage[utf8]{inputenc}
\usepackage[english]{babel}
\usepackage{amsmath,graphicx,enumerate}

\begin{document}

\title{Entropy production and collective excitations of crystals out of equilibrium: the concept of entropons} 


\author{L. Caprini$^{1}$}
\email{lorenzo.caprini@gssi.it, lorenzo.caprini@hhu.de}
\author{U. Marini Bettolo Marconi$^{2,3}$}
\author{H. L\"owen$^{1}$}

\affiliation{$^1$ Heinrich-Heine-Universit\"at D\"usseldorf, Universit\"atstrasse, 40225, D\"usseldorf, Germany.\\
$^2$ Scuola di Scienze e Tecnologie, Universit\`a di Camerino - via Madonna delle Carceri, 62032, Camerino, Italy.\\
$^3$ Istituto Nazionale di Fisica Nucleare, Sezione di Perugia, Via A. Pascoli, 06123 Perugia, Italy. 
}

\date{\today}


\begin{abstract}

We study the collective vibrational excitations of crystals under out-of-equilibrium steady conditions that give rise to entropy production. Their excitation spectrum comprises equilibrium-like phonons of thermal origin and additional collective excitations called entropons because each of them represents a mode of spectral entropy production.
Entropons coexist with phonons and dominate over them when the system is far from equilibrium while they are negligible in near-equilibrium regimes. The concept of entropons has been recently introduced and verified in a special case of crystals formed by self-propelled particles. Here, we show that entropons exist in a broader class of active cyrstals that are intrinsically out of equilibrium and characterized by the lack of detailed balance. 
After a general derivation, several explicit examples are discussed, including crystals consisting of particles with alignment interactions and frictional contact forces.

\end{abstract}

\maketitle


\section{Introduction}
Collective excitations play a fundamental role in the comprehension of solids and are one of the most fruitful concepts of solid-state physics~\cite{pines2018elementary}: prototypical examples are phonons, i.e.\ vibrational excitations of the elementary constituents of the crystal~\cite{chaikin1995principles}, and magnons,  the collective excitations associated with the electron spins in a crystal lattice~\cite{chaikin1995principles}.
In particular, in a crystal, the displacements of atoms from their equilibrium positions give rise to collective modes of vibrations named phonons whose amplitude is determined by the environmental temperature.
If the system is in equilibrium with its environment, there is no entropy production. In the framework of soft matter materials, it is possible to realize experimentally an equilibrium solid made of mesoscopic particles, instead of atoms, employing high-density colloidal suspensions and inducing its crystallization by decreasing the temperature and increasing the packing fraction~\cite{dinsmore1998self, li2016assembly}.
In equilibrium, the phase diagram of colloidal particles has been thoroughly investigated in the past for two and three-dimensional systems and different pairwise interaction potentials~\cite{lowen1994melting, hynninen2003phase, ivlev2012complex, hwang2019direct}, revealing a stable crystalline phase at high densities. The dynamics of these colloidal particles is Brownian overdamped motion as the surrounding solvent keeps the temperature of the system constant. 

Manipulations of solids can involve the use of external forces~\cite{schottle2022continuous}, such as laser pulses~\cite{kuzmany1970interaction}, acoustic fields~\cite{schottle2022continuous, menath2022acoustic}, and light fields~\cite{das2019active}.
External fields transfer energy to each particle of the solid and are responsible for entropy production~\cite{caprini2023ultrafast}, such that they can be considered one of the basic instances of out-of-equilibrium crystals~\cite{lowen2013introduction}.
Another important class of solids in non-equilibrium are active crystals. They consist of self-propelled agents that locally extract energy from the environment~\cite{marchetti2013hydrodynamics, bechinger2016active, gompper20202020} and convert it to perform specific tasks, such as directed motion.
For instance, solid structures are common at the micron scale in the realm of biology.
Examples are cell monolayers in human or animal bodies~\cite{alert2020physical, henkes2020dense, garcia2015physics}, biological tissues but also bacterial colonies at high density~\cite{petroff2015fast, peruani2012collective}.
Non-equilibrium crystals have been also investigated in active colloids, for instance by considering high-density active Janus particles that self-propel in space because of thermo- or electro-phoresis. They may form crystal structures with almost perfect hexagonal packing~\cite{buttinoni2013dynamical, van2019interrupted, ginot2018aggregation} that can even collectively travel or rotate~\cite{palacci2013living, mognetti2013living}.
%
%
%
Recently, solids made of out-of-equilibrium particles have been realised also at the macroscopic scale by using active granulars~\cite{baconnier2021selective}, i.e. granular particles dissipating energy through collisions that self-propel because of some shape asymmetry.

The investigation of collective excitations of non-equilibrium solids is now a challenging issue, relevant both to physics and biology, and requires linking together solid-state and non-equilibrium statistical physics concepts.
In the case of active crystals formed by self-propelled (active) particles, a novel kind of collective wavelike excitations has been discovered in Ref.~\cite{caprini2022entropons} (see Fig.~\ref{fig:Fig1}). These were named entropons because each of them is determined by the spectral entropy production of the system.
In the case investigated, entropons are sustained by the self-propelled (active) force acting and coexist without interfering with the usual thermal phonons.

In this paper, we show that entropons are not limited to the specific case of active solids considered in Ref.~\cite{caprini2022entropons}, that are formed by self-propelled particles, but are present in a broader class of non-equilibrium crystals that violate the detailed balance and produce entropy.
After considering a general set-up and giving a practical prescription to calculate the contributions of phonons and entropons,
we discuss a series of specific examples. 
These include crystals consisting of particles with alignment interactions and frictional contact forces.


The paper is structured as follows: in Sec.~\ref{sec:model}, we introduce a general model suitable to describe non-equilibrium solids, while in Sec.~\ref{sec:entropons}, we discuss the concept of entropons as non-equilibrium collective excitations coexisting with phonons, and provide a general prescription to calculate their contribution.
Section~\ref{Sec:general} contains a derivation of the main results of the previous section while Sec.~\ref{sec:examples} reports several examples where dynamical correlations and spectral entropy production are calculated analytically. Phonons and entropons are identified and discussed case by case.
Finally, we argue the consequences of our results and possible future research lines in the conclusions, Sec.~\ref{sec:conclusions}.

\section{Model}\label{sec:model}

We argue that the concept of entropons goes beyond the specific case of active solids studied in Ref.~\cite{caprini2022entropons}, 
by considering crystals formed by particles far from equilibrium but not necessarily self-propelled.
We require that
\begin{itemize}
\item[i)] The particles form a $d$-dimensional periodic lattice.
\item[ii)] Particles can only perform small fluctuations around their equilibrium positions so that their dynamics can be described in terms of displacement variables, $\mathbf{u}(t)$.
Due to the smallness of the fluctuations, the spatial Fourier transform of $\mathbf{u}(t)$ corresponding to a specific mode $\mathbf{q}(t)$ is decoupled from the remaining modes.
\item[iv)] The system reaches a (non-equilibrium) steady state in the long-time limit.
\end{itemize}
With these requirements, the particle-particle interactions can be easily treated and they determine the dispersion relation $\omega^2(\mathbf{q})$ within the harmonic approximation.

By taking advantage of the translational symmetry of a crystal, we can conveniently describe the system in Fourier space, in terms of frequency $\omega$ and wave vectors $\mathbf{q}$. 
By assuming a dissipative Brownian dynamics, subject to noise and dissipative (friction) forces, we can obtain an evolution equation for the Fourier transform of the particle displacement $\hat{\mathbf{u}}(\omega, \mathbf{q})$ at the frequency $\omega$ and wave vectors $\mathbf{q}$ (See Appendix~\ref{app:Fouriertransformdefinition} for definitions). 
This dynamics will be rather general and will include a broad range of equilibrium and non-equilibrium models usually studied in active matter and beyond.
Without loss of generality (see Sec.~\ref{sec:examples} for specific examples), the evolution equation for $\hat{\mathbf{u}}(\omega, \mathbf{q})$ can be expressed as
%
%
\begin{equation}
\label{eq:general_dynamics_u}
L(\omega, \mathbf{q}) \hat{\mathbf{u}}(\omega, \mathbf{q})=\hat{\mathbf{F}}(\omega, \mathbf{q}) + \sqrt{2T\gamma} \hat{\boldsymbol{\xi}}(\omega, \mathbf{q})
\end{equation}
where $\hat{\boldsymbol{\xi}}(\omega, \mathbf{q})$ is a white noise with zero average and correlation
\begin{equation}
\langle \hat{\boldsymbol{\xi}}(\omega, \mathbf{q})\cdot \hat{\boldsymbol{\xi}}(\omega, \mathbf{q}) \rangle =\delta(\omega+\omega') \delta(\mathbf{q}+\mathbf{q}')
\end{equation}
and the prefactor $T \gamma$ represents the amplitude of the thermal noise. Here, our nomenclature is inspired by the analogy with equilibrium solids in contact with a Brownian bath at temperature $T$ and embedded in a medium exerting a viscous friction of coefficient $\gamma$.
The term $L(\omega, \mathbf{q})$ is a complex function of $\omega$ and $\mathbf{q}$ (independent of the state variables, such as displacement, velocity, etc.) describing the evolution of the displacement $\hat{\mathbf{u}}(\omega, \mathbf{q})$. Note that with this formalism describes both overdamped and underdamped dynamics.
Here, for simplicity, we restrict ourselves to the case where $L(\omega, \mathbf{q})$ is a scalar operator acting equally on all spatial components and not a tensor, so that magnetic fields coupling different components are not considered.
Explicit examples for $L(\omega, \mathbf{q})$ (or its inverse) are provided in Sec.~\ref{sec:examples} both for equilibrium and non-equilibrium systems.

The term $\hat{\mathbf{F}}(\omega, \mathbf{q})$ is a force that does not depend on the particle displacement but can be a function of other dynamical variables involved in the system (see Sec.~\ref{sec:examples} for explicit examples). This force is zero in equilibrium conditions, while this term violates the detailed balanced and leads to entropy production in out-of-equilibrium conditions. To fix ideas, $\hat{\mathbf{F}}(\omega, \mathbf{q})$ may represent either the self-propelled force evolving through the dynamics of active particles or more complex dynamical stochastic processes that can even evolve non-linearly.
%

We remark that $L(\omega, \mathbf{q})$ can be conveniently decomposed onto its odd and even part under time-reversal transformation (TRT), $\omega\to-\omega$, according to:
\begin{equation}
L(\omega, \mathbf{q})=L_{o}(\omega, \mathbf{q})+L_{e}(\omega, \mathbf{q})
\end{equation}
where the subscripts $o$ and $e$ mean odd and even, respectively, under TRT, so that
$L_{o} \to - L_{o}$ and $L_{e} \to L_{e}$.
As intuition suggests, $\hat{\mathbf{u}}(\omega, \mathbf{q})$ is even under TRT and, for simplicity, we restrict our analysis to a set of dynamical variables $\mathbf{F}(\omega, \mathbf{q})$ that are even under TRT so that $\mathbf{F} \to \mathbf{F}$.

\section{The concept of entropons}\label{sec:entropons}

\begin{figure}[!t]
\centering
\includegraphics[width=1\linewidth,keepaspectratio] {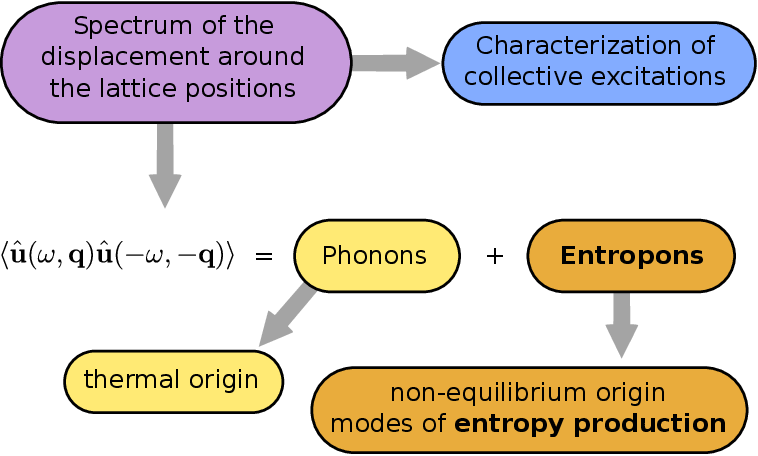}
\caption{{\textbf{Collective excitations in non-equilibrium crystals.}} By analyzing the spectrum of the particles' displacement around the lattice positions, it is possible to characterize the collective vibrations of crystals.
Crystals out of equilibrium are characterized by phonons and additional novel collective excitations that we called entropons because they are generated by entropy production.
}
\label{fig:Fig1}
\end{figure}

In this section, we anticipate our results by introducing the concept of entropons as collective excitations which originate from non-equilibrium. Here, the meaning of entropons is discussed, while the derivation of our results is reported in Sec.~\ref{Sec:general}.

To characterize collective excitations in non-equilibrium solids, we study the dynamical correlations of the Fourier transform of the particle displacements around their lattice positions, $\mathcal{C}(\omega, \mathbf{q})$, defined in the Fourier space of frequency $\omega$ and wave vector $\mathbf{q}$, as
\begin{equation}
\label{eq:Comegaq_def}
\mathcal{C}(\omega, \mathbf{q}) = \lim_{\mathcal{T}\to\infty} \frac{1}{\mathcal{T}} \langle \hat{\mathbf{u}}(\omega, \mathbf{q}) \cdot \hat{\mathbf{u}}(-\omega, -\mathbf{q})\rangle \,.
\end{equation}
The dynamical correlations $\mathcal{C}(\omega, \mathbf{q})$ can be conveniently decomposed as
\begin{equation}
\label{eq:definitionEntropons}
\mathcal{C}(\omega, \mathbf{q}) = \mathcal{C}_{\text{eq}}(\omega, \mathbf{q}) + \mathcal{C}_{\text{out}}(\omega, \mathbf{q})
\end{equation}
where $\mathcal{C}_{eq}(\omega, \mathbf{q})$ and $\mathcal{C}_{out}(\omega, \mathbf{q})$ are the equilibrium and out-of-equilibrium parts of the dynamical correlation of the particle displacement $\hat{\mathbf{u}}(\omega, \mathbf{q})$, respectively. As clarified later, the first part has a thermal origin, while the second part originates from the non-equilibrium force pushing the system out of equilibrium.
$\mathcal{C}_{eq}(\omega, \mathbf{q})$ can be expressed in terms of the response function to a small perturbation while $\mathcal{C}_{out}(\omega, \mathbf{q})$ can be related to the spectral entropy production of the system $\sigma(\omega, \mathbf{q})$.
As a consequence, the decomposition~\eqref{eq:Comegaq_def} can be interpreted as a generalization of the Harada-Sasa relation~\cite{harada2005equality} for the case of nonequilibrium solids.

As obtained in Sec.~\ref{Sec:general}, $\mathcal{C}_{eq}(\omega, \mathbf{q})$ can be written as
\begin{equation}
\label{eq:Cequilibrium}
\mathcal{C}_{eq}(\omega, \mathbf{q})= - 2T\gamma \frac{\text{Im}[\mathcal{R}_{uu}(\omega, \mathbf{q})]}{\text{Im}[L(\omega, \mathbf{q})]}
\end{equation}
where $\text{Im}[\cdot]$ denotes the imaginary part and $\mathcal{R}_{uu}(\omega, \mathbf{q})$ is the Fourier transform of the displacement response function due to a small perturbation, $\mathbf{h}$, defined as
\begin{equation}
\label{eq:responsedefinition}
\mathcal{R}_{\hat{u}\hat{u}}(\omega, \mathbf{q}) = \text{Tr}\left[ \frac{\delta \langle \mathbf{u}(\omega, \mathbf{q})\rangle_h}{\delta \mathbf{h}}\right] \,.
\end{equation}
Here, $\text{Tr}[\cdot]$ stands for the trace of the matrix inside the square brackets.
The average $\langle \cdot\rangle_h$ is defined over the perturbed trajectory, and $\delta/\delta \mathbf{h}$ is the functional derivative with respect to the perturbation $\mathbf{h}$, as usual in linear response theory \cite{crisanti2003violation, marconi2008fluctuation, cugliandolo2011effective, maes2020frenesy}.
As known in the literature, $\mathcal{R}_{\hat{u}\hat{u}}(\omega, \mathbf{q})$ can be explicitly calculated and in our linear model we find
\begin{equation}
\mathcal{R}_{\hat{u}\hat{u}}(\omega, \mathbf{q}) = L^{-1}(\omega, \mathbf{q}) \,.
\label{eq:seven}
\end{equation}
From Eq. \eqref{eq:general_dynamics_u} it is clear that $\mathcal{R}_{\hat{u}\hat{u}}(\omega, \mathbf{q})$ is independent of the non-equilibrium force $\mathbf{F}(\omega, \mathbf{q})$ and we may anticipate that it is associated with the phononic spectrum.

The contribution $\mathcal{C}_{out}(\omega, \mathbf{q})$ can be related to the entropy production of the system, that quantifies how the system is far from equilibrium.
As derived in Sec.~\ref{Sec:general}, the expression for $\mathcal{C}_{out}(\omega, \mathbf{q})$ can be explicitly calculated as
\begin{equation}
\label{eq:Cnonequilibrium}
\frac{\mathcal{C}_{out}(\omega, \mathbf{q})}{T\gamma}
= \frac{\sigma(\omega, \mathbf{q})}{(\text{Im}[L(\omega, \mathbf{q})])^2}
\end{equation}
where $\sigma(\omega, \mathbf{q})$, is the spectral entropy production, i.e. the spectral component (in frequency and wave vector domains) of the total entropy production rate, $\dot{s}$, such that
\begin{equation}
\label{eq:defspectralEP}
\dot{s}= \int \frac{d\mathbf{q}}{\Omega} \int \frac{d\omega}{2\pi}\, \sigma(\omega, \mathbf{q}) \,.
\end{equation}
Here, $\Omega$ represents the volume of the first Brillouin zone, depending on the lattice properties of the solid.
As shown later, $\mathcal{C}_{out}(\omega, \mathbf{q})$ represents additional collective excitations of the system that we identify as entropons.

For the general dynamics~\eqref{eq:general_dynamics_u}, $\sigma(\omega, \mathbf{q})$ can be calculated using a path-integral method, in frequency and wave vector domains (see Sec.~\ref{Sec:general}), and is given by
\begin{equation}
\label{eq:explicit_sigma_expression}
\sigma(\omega, \mathbf{q})= \lim_{t\to\infty} \frac{i}{t}\frac{ \text{Im}[L(\omega, \mathbf{q})] }{2 T \gamma}\langle \hat{\mathbf{u}}(\omega, {\mathbf{q}})\hat{\mathbf{F}}(-\omega, -{\mathbf{q}})\rangle + c.c
\end{equation}
where $c.c.$ denotes the complex conjugate. As a consequence, $\sigma(\omega, \mathbf{q})$ is real and requires only the knowledge of $L(\omega, \mathbf{q})$ and the cross-correlation $\langle \hat{\mathbf{u}}(\omega, \mathbf{q}) \cdot\hat{\mathbf{F}}(-\omega, -\mathbf{q})\rangle$.

\subsection{Coexistence of phonons and entropons}

Here, we present the physical interpretation of the decomposition~\eqref{eq:definitionEntropons} together with Eqs.~\eqref{eq:Cequilibrium} and~\eqref{eq:Cnonequilibrium}.
As schematically illustrated in Fig.~\ref{fig:Fig1}, relation~\eqref{eq:definitionEntropons} states that the non-equilibrium excitations of a solid, described by the dynamical correlation of the particle displacement, can be decomposed in two parts: i) an equilibrium-like contribution $\mathcal{C}_{eq}(\omega, \mathbf{q})$ entirely due to the thermal noise (phonons), ii) a non-equilibrium contribution $\mathcal{C}_{out}(\omega, \mathbf{q})$ proportional to the spectral entropy production of the system (entropons).

{\it \bf Phonons.} Term i) has the same form as the displacement-displacement dynamical correlation of an equilibrium
underdamped solid consisting of particles in contact with a thermal bath.
It describes the thermally excited collective vibrations of crystals, i.e. the familiar phonons typical of solid-state physics: for a given $\mathbf{q}$, a peak in the profile of $\mathcal{C}_{eq}(\omega, \mathbf{q})$ as a function of $\omega$ can be identified with a phonon of the crystal. Equation~\eqref{eq:definitionEntropons} suggests that phonons are present both in equilibrium and non-equilibrium solids, their spectrum remains unaltered and they do not generate entropy production. The non-equilibrium force does not affect their dispersion.

{\it \bf Entropons.}
Term ii), $\mathcal{C}_{out}(\omega, \mathbf{q})$, describes new vibrational collective excitations of the crystal of truly non-equilibrium origin as $\mathcal{C}_{out}(\omega, \mathbf{q})$ is proportional to the spectral entropy production of the system $\sigma(\omega, \mathbf{q})$.
Indeed, entropons vanish at equilibrium together with $\sigma(\omega, \mathbf{q})$.
As typical in solid-state physics, the peaks of $\mathcal{C}_{out}(\omega, \mathbf{q})$ as a function of $\omega$ (at fixed $\mathbf{q}$) are identified with these excitations.
We term them entropons because for each value of $\mathbf{q}$ there is a component of the spectral entropy production, $\sigma(\omega, \mathbf{q})$.
Entropons coexist with phonons and remain distinct from them.
At a fixed $\mathbf{q}$ the frequencies corresponding to their peaks differ from those of phonons.
The amplitude of entropons is negligible with respect to that of phonons in near-equilibrium conditions, where the entropy production is small, whereas far from the equilibrium entropons play the dominant role.
Entropons will be shown and discussed more specifically through explicit examples in Sec.~\ref{sec:examples}.

\section{Derivation of the result}\label{Sec:general}

We now prove the decomposition~\eqref{eq:definitionEntropons} and formulas~\eqref{eq:Cequilibrium} and~\eqref{eq:Cnonequilibrium},
and take advantage of the linearity of the system to derive analytically the correlation $\mathcal{C}(\omega, \mathbf{q})$, the response function $\mathcal{R}_{uu}(\omega, \mathbf{q})$ and the entropy production $\sigma(\omega, \mathbf{q})$. Finally, $\mathcal{C}_{eq}(\omega, \mathbf{q})$ and $\mathcal{C}_{out}(\omega, \mathbf{q})$ are identified.

\subsection{Dynamical correlations of the displacements}

To derive the analytical expression for the dynamical correlations of the particle displacements, it is convenient to introduce the notation ${G}(\omega, \mathbf{q})=L^{-1}(\omega, \mathbf{q})$ as the inverse of $L(\omega, \mathbf{q})$.
From the linearity of the model, the solution for each Cartesian component of the displacement $\hat{u}(\omega, \mathbf{q})$ (for instance the $x$ component) is given by
\begin{equation}
\label{eq:solutionforu}
\hat{u}(\omega, \mathbf{q}) = G(\omega, \mathbf{q}) \hat{F}(\omega, \mathbf{q}) + G(\omega, \mathbf{q}) \sqrt{2T\gamma} \hat{\xi}(\omega, \mathbf{q}) \,.
\end{equation}
By multiplying Eq.~\eqref{eq:solutionforu} by $\hat{u}(-\omega, -\mathbf{q})$ and averaging over the noise, we get
\begin{equation}
\begin{aligned}
&\langle \hat{u}(\omega, \mathbf{q})\hat{u}(-\omega, -\mathbf{q})\rangle =\\
&+(2T\gamma) G(\omega, \mathbf{q}) G(-\omega, -\mathbf{q})\langle \hat{\xi}(\omega, \mathbf{q}) \hat{\xi}(-\omega, -\mathbf{q}) \rangle  \\
&+G(\omega, \mathbf{q}) G(-\omega, -\mathbf{q})\langle \hat{F}(\omega, \mathbf{q})\hat{F}(-\omega, -\mathbf{q}) \rangle \,,
\end{aligned}
\end{equation}
while by accounting for Eq.~\eqref{eq:Comegaq_def}, we obtain
\begin{flalign}
&\mathcal{C}(\omega, \mathbf{q})=
(2T\gamma) G(\omega, \mathbf{q}) G(-\omega, -\mathbf{q})\\
&+\lim_{t\to\infty} \frac{1}{t} G(\omega, \mathbf{q}) \cdot \langle \hat{F}(\omega, \mathbf{q})\hat{F}(-\omega, -\mathbf{q}) \rangle G(-\omega, -\mathbf{q}) \nonumber\,.
\end{flalign}
Quite intuitively, we identify the equilibrium and non-equilibrium parts of the dynamical correlations as
\begin{flalign}
\label{eq:Ceqexpression}
&\mathcal{C}_{eq}(\omega, \mathbf{q}) =(2T\gamma) G(\omega, \mathbf{q}) G(-\omega, -\mathbf{q})\\
\label{eq:Cnoneqexpression}
&\mathcal{C}_{out}(\omega, \mathbf{q}) =\lim_{t\to\infty} \frac{1}{t} G(\omega, \mathbf{q}) \langle \hat{F}(\omega, \mathbf{q})\hat{F}(-\omega, -\mathbf{q}) \rangle G(-\omega, -\mathbf{q}) \,.
\end{flalign}
The first line corresponds to the effect of the thermal noise while the second line to the one of the non-equilibrium force.
We remark that the above results are obtained without specifying the dynamics of $\hat{F}(\omega, \mathbf{q})$ and could be valid under more general conditions, even in the presence of non-linearities. However, a non-linear evolution equation for $\hat{F}(\omega, \mathbf{q})$, renders much harder or even impossible the analytic determination of the correlation function $\langle \hat{F}(\omega, \mathbf{q})\hat{F}(-\omega, -\mathbf{q}) \rangle$.

\subsection{Response function}

By adding a small perturbative force $ h(\omega, \mathbf{q})$ to Eq.~\eqref{eq:solutionforu} the resulting perturbed dynamics
reads:
\begin{equation}
L(\omega, \mathbf{q}) \hat{u}(\omega, \mathbf{q})=\hat{F}(\omega, \mathbf{q})  + \sqrt{2T\gamma} \hat{\xi}(\omega, \mathbf{q}) +  h(\omega, \mathbf{q})
\end{equation}
and
applying the definition~\eqref{eq:responsedefinition}, we derive with respect to $h(\omega,\mathbf{q})$ and obtain the response
$
\mathcal{R}_{\hat{u}\hat{u}}(\omega, \mathbf{q}) = G(\omega, \mathbf{q}) \,$
which coincides with \eqref{eq:seven}.
We remark that in virtue of the linearity of the system $\mathcal{R}_{\hat{u}\hat{u}}(\omega, \mathbf{q})$ is not affected by $\hat{F}(\omega, \mathbf{q})$, i.e. the dynamical variables pushing the system out of equilibrium.
By this identification, the equilibrium part of the correlation, $\mathcal{C}_{eq}(\omega, \mathbf{q})$, defined in Eq.~\eqref{eq:Ceqexpression}, can be rewritten as
\begin{equation}
\mathcal{C}_{eq}(\omega, \mathbf{q})= 2T\gamma \mathcal{R}_{uu}(\omega, \mathbf{q}) \mathcal{R}_{uu}(-\omega, -\mathbf{q}) \,.
\end{equation}
Alternatively, by using the properties of the complex numbers, and, in particular, the general relation
\begin{equation}
\label{eq:relationcomplexnumber}
|G(\omega, \mathbf{q})|^2 =- \frac{\text{Im}[G(\omega, \mathbf{q})]}{\text{Im}[G^{-1}(\omega, \mathbf{q})]}
\end{equation}
 we can express $\mathcal{C}_{eq}(\omega, \mathbf{q})$ in a more familiar form as
\begin{equation}
\label{eq:der_Ceq}
\frac{\mathcal{C}_{eq}(\omega, \mathbf{q})}{2T\gamma}
= - \frac{\text{Im}[\mathcal{R}_{uu}(\omega, \mathbf{q})] }{\text{Im}[L(\omega, \mathbf{q})]}
\end{equation}
that corresponds to Eq.~\eqref{eq:Cequilibrium}, i.e.\ the contribution of phonons to the correlation function.
Note that in the specific case for which the Harada-Sasa relation has been proposed, we have ${\text{Im}[L(\omega, \mathbf{q})]}=\omega$~\cite{harada2005equality}.

\subsection{Calculation of the spectral entropy production}

The spectral entropy production $\sigma(\omega, \mathbf{q})$ can be operatively calculated by using path-integral methods in frequency $\omega$ and wave vector $\mathbf{q}$ domains.
In the framework of stochastic thermodynamics~\cite{seifert2012stochastic, speck2016stochastic, szamel2019stochastic}, the entropy production $\dot{s}$ measures the degree of irreversibility of the trajectory of a stochastic system~\cite{o2022time} and is defined through path-integral methods as~\cite{seifert2012stochastic, spinney2012nonequilibrium, pigolotti2017generic, caprini2019entropy, dabelow2019irreversibility}
\begin{equation}
\label{eq:entropy_prod}
\dot{s} = \lim_{t\to\infty}\frac{1}{t}\left\langle\log{\left[\frac{P(\{\hat{u}\}|\hat{u}_0)}{P_r(\{\hat{u}\}|\hat{u}_0)}\right]}\right\rangle
\end{equation}
where $P(\{\hat{u}\}|\hat{u}_0)$ and $P_r(\{\hat{u}\}|\hat{u}_0)$ are the probability of forward and backward trajectories, respectively.
The path probabilities depend on the whole time history of the dynamical variables (denoted by curly brackets $\{ \cdot\}$) and are conditioned to the initial value $\hat{u}_0$.
From now, we denote variables or observables of the reverse dynamics with the subscript $r$.
The easier way to derive $P$ and $P_r$ is starting from the probability distribution of the Gaussian noise vector $\hat{w}$ (forward) and $\hat{\xi}_r$ (backward), conditioned to the initial value $\hat{\xi}_0$, and given by
\begin{subequations}
\label{eq:prob_traj}
\begin{align}
&\text{p}(\{\hat{\xi}\}|\hat{\xi}_0)\sim e^{\mathcal{A}}\\
&\text{p}_r(\{\hat{\xi}\}|\hat{\xi}_0) \sim e^{\mathcal{A}_r} \,.
\end{align}
\end{subequations}
where $\mathcal{A}$ and $\mathcal{A}_r$ read
\begin{subequations}
\label{eq:action}
\begin{align}
&\mathcal{A} = -\frac{1}{2}\int \frac{d\omega}{2\pi} \sum_{\mathbf{q}} \hat{\xi}(\omega, {\mathbf{q}}) \hat{\xi}(-\omega, -{\mathbf{q}}) \,,\\
&\mathcal{A}_r = -\frac{1}{2}\int \frac{d\omega}{2\pi} \sum_{\mathbf{q}} \hat{\xi}_r(\omega, {\mathbf{q}}) \hat{\xi}_r(-\omega, -{\mathbf{q}}) \,.
\end{align}
\end{subequations}
From here, we identify $p\sim P$ and $p_r \sim P_r$, by performing a change of variables $\hat{\xi}(\omega, \mathbf{q}) \to \hat{u}(\omega, \mathbf{q})$, using the equation of motion~\eqref{eq:solutionforu}.
To carry out this program, one should estimate the determinant of the transformation but, as known, in the additive-noise case, the determinant does not affect the expression for the entropy production and can be safely ignored~\cite{spinney2012nonequilibrium, spinney2012entropy}.
In practice, $\mathcal{A}$ and $\mathcal{A}_r$ are identified as the forward and backward actions associated with the dynamics by replacing
\begin{equation}
\label{eq:xi_forward}
\hat{\xi}(\omega, \mathbf{q})= \frac{(L_{e}(\omega, \mathbf{q})+L_{o}(\omega, \mathbf{q})) \hat{u}(\omega, \mathbf{q}) - \hat{F}(\omega, \mathbf{q})}{\sqrt{2T\gamma} }
\end{equation}
and
\begin{equation}
\label{eq:xi_backward}
\hat{\xi}_r(\omega, \mathbf{q})= \frac{(L_{e}(\omega, \mathbf{q})-L_{o}(\omega, \mathbf{q})) \hat{u}(\omega, \mathbf{q}) - \hat{F}(\omega, \mathbf{q})}{\sqrt{2T\gamma} }
\end{equation}
where the expression of $\hat{\xi}_r$ is obtained by applying the time-reversal transformation to the dynamics~\eqref{eq:solutionforu}, i.e. using that $\omega_r =-\omega$, $\hat{u}_r=\hat{u}$ and $\hat{F}_r=\hat{F}$ because according to our choice, also $F$ is even under time-reversal transformation.

By using the definition~\eqref{eq:defspectralEP}, one can identify the spectral entropy production as
\begin{equation}
\begin{aligned}
\sigma(\omega, \mathbf{q})=& \lim_{t\to\infty} \frac{1}{t} \frac{\langle \hat{\xi}_r(\omega, {\mathbf{q}}) \hat{\xi}_r(-\omega, -{\mathbf{q}})\rangle}{2}\\
&- \lim_{t\to\infty} \frac{1}{t}\frac{\langle \hat{\xi}(\omega, {\mathbf{q}}) \hat{\xi}(-\omega, -{\mathbf{q}})\rangle}{2}\,.
\end{aligned}
\end{equation}
After standard algebraic manipulations, obtained by using Eq.~\eqref{eq:xi_forward} and Eq.~\eqref{eq:xi_backward}, $\sigma(\omega, \mathbf{q})$ reads
\begin{equation}
\label{eq:spectralEntropyprod_derivation}
\sigma(\omega, \mathbf{q})=\lim_{t\to\infty} \frac{1}{t} \frac{L_{o}(\omega, \mathbf{q}) \langle \hat{u}(\omega, {\mathbf{q}}) \hat{F}(-\omega, -{\mathbf{q}})\rangle}{2 T \gamma} + c.c.
\end{equation}
Recalling that $L_o(\omega, \mathbf{q})=i\text{Im}[L(\omega, \mathbf{q})]$, we immediately obtain the explicit expression for $\sigma(\omega, \mathbf{q})$, Eq.~\eqref{eq:explicit_sigma_expression}.

Plugging Eq.\eqref{eq:relationcomplexnumber} into the expression for $\sigma(\omega, \mathbf{q})$ and using $\langle \hat{\xi}\rangle=0$, we obtain
\begin{flalign}
\label{eq:sigma_usefulformula}
&\sigma(\omega, \mathbf{q})= \lim_{t\to\infty} \frac{1}{t}\frac{\text{Re}[L_{o}(\omega, \mathbf{q}) \langle \hat{u}(\omega, {\mathbf{q}}) \hat{F}(-\omega, -{\mathbf{q}})\rangle]}{T \gamma}\\
&=\lim_{t\to\infty} \frac{1}{t}\frac{\langle \hat{F}(\omega, {\mathbf{q}}) \hat{F}(-\omega, -{\mathbf{q}})\rangle \text{Re}[L_{o}(\omega, \mathbf{q}) G(\omega, \mathbf{q})]}{T \gamma}\nonumber\\
&= - \lim_{t\to\infty} \frac{1}{t} \frac{\text{Im}[L(\omega, \mathbf{q})] \text{Im}[G(\omega, \mathbf{q})] \langle \hat{F}(\omega, {\mathbf{q}}) \hat{F}(-\omega, -{\mathbf{q}})\rangle}{T \gamma}\nonumber
\end{flalign}
where $\text{Re}[\cdot]$ means real part and, in the last equality, we have used the properties $L_{o}(\omega, \mathbf{q})=i \,\text{Im}[L(\omega,\mathbf{q})]$ due to the linearity of $L$.
Finally, by using Eq.~\eqref{eq:relationcomplexnumber} to replace $\text{Im}[G(\omega, \mathbf{q})]$ and the expression for $\mathcal{C}_{out}(\omega, \mathbf{q})$ (Eq.~\eqref{eq:Cnoneqexpression}), we have
\begin{equation}
\sigma(\omega, \mathbf{q})= \frac{\mathcal{C}_{out}(\omega, \mathbf{q}) [\text{Im}{L}(\omega, \mathbf{q})]^{2}}{T \gamma}
\end{equation}
that coincides with Eq.~\eqref{eq:Cnonequilibrium} and concludes the derivation of our results for the general dynamics~\eqref{eq:solutionforu}.


 
\section{Examples}\label{sec:examples}

In this section, we report several explicit examples of solids, formed by particles in contact with a thermal bath and described by underdamped equations of motion, for their positions, $\mathbf{x}_i$, and the velocities, $\mathbf{v}_i$.
They interact through the total pairwise potential, $U_{tot}$, given by
\begin{equation}
U_{tot} =\sum_{i<j}^N U(|\mathbf{x}_j -\mathbf{x}_i|)
\end{equation}
where $U(r)$ is a generic interaction potential that only depends  on the distance $r$ between a pair of particles.
The present theory holds for general dimensionality and general potentials, which can be attractive or repulsive, short- or long-range. In all cases, interactions must be such that particles arrange in solid-like configurations in a typical lattice structure where defects are not statistically relevant and can be neglected. The system should be characterized by large values of the density close to the packing regime and/or values of equilibrium and non-equilibrium fluctuations so that the crystalline phase is maintained.

For the sake of simplicity, here, we restrict our discussion to the case of short-range forces so that a particle interacts only with its first neighbors.
Under this assumption, we Taylor expand the potential around its minimum and obtain
\begin{equation}
\label{eq:Utot}
U_{tot} \approx \frac{m\omega_E^2}{2} \sum_{i<j}^{*} \left(\mathbf{u_i}-\mathbf{u_j}\right)^2
\end{equation}
where $\mathbf{u}_j$ is the displacement of the particle $j$ from its lattice position and the sum $\overset{*}{\sum}$ is restricted only to first neighbors.
The quantity $\omega_E$ represents the Einstein frequency of the solid and depends on the spatial second derivatives of $U(r)$ evaluated at the lattice constant. Its functional form is determined by the dimensions and the structure of the lattice.
Explicit expressions of $\omega_E$ are reported in Appendix~\ref{app:dispersionrelation}.

In virtue of the approximations performed, the force acting on each particle of the solid, $\mathbf{F}^{lattice}_i$, can be approximated as
\begin{equation}
\label{eq:linearforceapprox}
\mathbf{F}^{lattice}_i\approx - m\omega^2_E \sum_j^* \left(\mathbf{u}_i - \mathbf{u}_j \right) \,.
\end{equation}
Its Fourier transform in the domains of frequency, $\omega$, and wave vectors, $\mathbf{q}$, reads
\begin{equation}
\hat{\mathbf{F}}^{lattice}(\omega, \mathbf{q}) = - m\omega^2(\mathbf{q}) \hat{\mathbf{u}}(\omega, \mathbf{q}) \,,
\end{equation}
where $\omega^2(\mathbf{q})\propto \omega_E^2/m$, is the dispersion relation determined by the geometry of the lattice structure and the interaction.
Explicit examples for $\omega^2(\mathbf{q})$ are reported in Appendix~\ref{app:dispersionrelation}.
In this description, the interaction force is accounted for in the term $L(\omega,\mathbf{q}) \hat{\mathbf{u}}(\omega, \mathbf{q})$ of Eq.~\eqref{eq:general_dynamics_u}.

\subsection{Equilibrium crystals}

In the framework of soft materials, equilibrium crystals are periodic structures consisting of particles in equilibrium with the environment.
Examples are passive colloidal systems at high density, for which inertia is really small and usually neglected, and complex plasma, described by an underdamped equation of motion where the degree of damping can be even steered~\cite{ivlev2012complex}.

A crystal formed by particles in equilibrium with a thermal bath at temperature $T$ is described by the following underdamped dynamics:
\begin{subequations}
\label{eq:equilibriumdynamics}
\begin{align}
&\dot{\mathbf{x}}_i=\mathbf{v}_i\\
&m\dot{\mathbf{v}}_i= -\gamma \mathbf{v}_i + \mathbf{F}^{lattice}_i +\sqrt{2 T \gamma} \boldsymbol{\xi}_i
\end{align}
\end{subequations}
where $\boldsymbol{\xi}_i$ are vectors of white noise with zero average and such that $\langle \boldsymbol{\xi}_i(t) \boldsymbol{\xi}_i(0)\rangle =\delta_{ij}\delta(t)$.
The energy injected by the thermal noise, $\sqrt{2 T \gamma} \boldsymbol{\xi}_i$, is dissipated in the environment through the viscous force $-\gamma \mathbf{v}_i$, proportional to the friction coefficient $\gamma$.
The force between particles that guarantees the solid structure (large density regime) $\mathbf{F}^{lattice}_i$ is given by Eq.~\eqref{eq:linearforceapprox}.
In this system, we can identify the ratio $\tau_I=m/\gamma$ as the inertial time of the system, i.e. the time necessary for the velocity to relax under the influence of the linear friction force.

The dynamics in Fourier space reads:
\begin{equation}
\label{eq:dynamics_thermalsolid}
\left(- m \omega^2 + i\omega\gamma +m \omega^2(\mathbf{q}) \right)\hat{\mathbf{u}}(\omega, \mathbf{q})=\sqrt{2T\gamma} \hat{\boldsymbol{\xi}}(\omega, \mathbf{q})
\end{equation}
where the hat-symbol denotes the double $\omega,\mathbf{q}$ Fourier transform. We also recall that the Fourier transform of the velocity is related to the displacement by 
$i\omega \hat{\mathbf{u}}(\omega, \mathbf{q})=\hat{\mathbf{v}}(\omega, \mathbf{q})$ and that the Fourier transform of a white noise with zero average, i.e. $\hat{\boldsymbol{\xi}}(\omega, \mathbf{q})$, has zero average and correlation $\langle \hat{\boldsymbol{\xi}}(\omega, \mathbf{q})\hat{\boldsymbol{\xi}}(\omega', \mathbf{q}')\rangle=\delta(\omega+\omega') \delta(\mathbf{q}+\mathbf{q}')$.

Applying the general methods, described in Sec.~\ref{sec:entropons}, the dynamical correlations in Fourier space can be analytically calculated after identifying the operator $G(\omega, \mathbf{q})=L^{-1}(\omega, \mathbf{q})$ as
\begin{equation}
\label{eq:Gthermal}
G(\omega, \mathbf{q}) = \frac{1}{ m \omega^2(\mathbf{q})- m \omega^2 + i \omega \gamma}
\end{equation}
and consequently, $L(\omega, \mathbf{q})$ as its inverse.
By applying Eq.~\eqref{eq:Cequilibrium}, the equilibrium dynamical correlation $\mathcal{C}_{eq}(\omega,\mathbf{q})$ yields
\begin{equation}
\label{eq:Ceq_thermal}
\gamma\frac{\mathcal{C}_{eq}(\omega, \mathbf{q})}{T}
= \frac{2}{\tau_I^2(\omega^2(\mathbf{q})-\omega^2)^2 +\omega^2} \,
\end{equation}
while Eq.~\eqref{eq:explicit_sigma_expression} implies that
\begin{equation}
\sigma(\omega, \mathbf{q})=0 \,.
\end{equation}
The system does not produce entropy, and, as a consequence, entropons disappear.
This is the expected result in the case of equilibrium-like solids, for which the dynamical correlations are pure phonons.
The frequency spectrum is, of course, affected by $\tau_I$, which has to be compared with the Einstein frequency $\omega_E$: for $\tau_I\omega_E \gg 1$ (small damping regime) converges to a Dirac $\delta$-function displaying a peak at $\omega \sim \omega(\mathbf{q})$, while for $\tau_I\omega_E\to0$ its shape flattens.
Phonons are excited by thermal fluctuations and, indeed, disappear in the limit $T\to0$ in the absence of Brownian fluctuations.

\subsection{Self-propelled solids}

Active systems are characterized by an internal mechanism, often represented as an additional degree of freedom, that converts energy from the environment to produce directed (self-propelled) motion~\cite{elgeti2015physics, bechinger2016active}.
Coarse-grained non-equilibrium stochastic models are widely employed in the theoretical descriptions of active particles both in overdamped and underdamped regimes: a popular approach 
consists in adding a time-dependent stochastic force, $\mathbf{f}^a_i$, to the velocity dynamics Eq.~\eqref{eq:equilibriumdynamics}. This force is a convenient representation of the self-propulsion mechanism, which is a chemical reaction in the case of Janus particles or the movement of flagella in the case of bacteria, for instance. The self-propulsion $\mathbf{f}^a_i$ is, in general, responsible for the persistent trajectories experimentally observed in these systems.
The resulting equation of motion reads
\begin{subequations}
\label{eq:equilibriumdynamicsc}
\begin{align}
&\dot{\mathbf{x}}_i=\mathbf{v}_i\\
&m\dot{\mathbf{v}}_i= -\gamma \mathbf{v}_i + \mathbf{F}^{lattice}_i +\sqrt{2 T \gamma} \boldsymbol{\xi}_i + \mathbf{f}^a_i \,.
\end{align}
\end{subequations}
The self-propelled (or active) force $\mathbf{f}^a_i$ endows the particle  with a swim velocity, $v_0$,  and takes the form
\begin{equation}
\mathbf{f}^a_i =\gamma v_0 \mathbf{n}_i
\end{equation}
where $\mathbf{n}_i$ is a stochastic process whose dynamics depends of the specific model under consideration.

Within the active Brownian particle (ABP) model~\cite{fily2012athermal, solon2015pressure, sese2018velocity, caprini2020spontaneous, caporusso2020motility, breoni2020active, hecht2022active, caraglio2022analytic}, $\mathbf{n}_i$ is represented as a unit vector,  $\mathbf{n}_i=(\cos\theta_i, \sin\theta_i)$, where $\theta_i$ represents the orientational angle of the active particle, evolving as
\begin{equation}
\dot{\theta}_i = \sqrt{2D_r}\eta_i \,.
\end{equation}
Here, $\eta_i$ is a white noise vector with unit variance and zero average and the prefactor sets the value of the rotational diffusion coefficient $D_r$. It also determines the persistence time of the particle trajectory, $\tau=1/D_r$ (in two dimensions), i.e. the time after which the orientation of the active force is randomized~\cite{farage2015effective, caprini2022parental}.

Recently, the active Ornstein-Uhlenbeck particle (AOUP) model~\cite{szamel2014self, wittmann2018effective, caprini2019activityinduced, woillez2020nonlocal, martin2021statistical, PhysRevLett.129.048002} has been proposed as an alternative to ABP.
AOUP was originally introduced to describe the behavior of a passive tracer in a non-equilibrium bath of active particles (bacteria, precisely)~\cite{wu2000particle, maggi2014generalized}, and later has been used as a theoretical simplification of the ABP~\cite{fily2012athermal, caprini2021spatial}.
According to the AOUP, $\mathbf{n}_i$ evolves as an Ornstein-Uhlenbeck process with typical time $\tau$ and unit variance
\begin{equation}
\tau\dot{\mathbf{n}}_i = - \mathbf{n} + \sqrt{2\tau} \boldsymbol{\eta}_i \,,
\end{equation}
where $\boldsymbol{\eta}_i$ is a vector of white noises with zero average and unit variance.
AOUPs show similar phenomena compared to ABPs, displaying accumulation near walls~\cite{caprini2018active, das2018confined, caprini2019activechiral} and collective phenomena, such as motility induced phase separation~\cite{fodor2016far, maggi2021universality}, and non-equilibrium spatial velocity correlations~\cite{caprini2020spontaneous, caprini2020hidden, flenner2016nonequilibrium, szamel2021long, PhysRevResearch.5.013077} in dense active systems where the AOUP theory has been employed to interpret the results from ABP simulations, for instance predicting the value of the kinetic temperature~\cite{caprini2020activekinetic}.
Further details concerning the relation between the two models are provided in Ref.~\cite{caprini2022parental}.

This energy exchange induced by the active force pushes a self-propelled particle out-of-equilibrium and leads to entropy production~\cite{fodor2016far, pietzonka2017entropy, grandpre2021entropy, dabelow2021irreversible, puglisi2017clausius, caprini2019entropy, mandal2017entropy, caprini2018comment, dadhichi2018origins}, even in the absence of external forces~\cite{shankar2018hidden}.
Except for special cases~\cite{cocconi2020entropy, razin2020entropy, frydel2022intuitive}, such as potential-free particles~\cite{shankar2018hidden, chaki2018entropy} and harmonic confinement~\cite{garcia2021run}, entropy production in active systems can be investigated only numerically, for instance in active field theories~\cite{nardini2017entropy, borthne2020time, paoluzzi2022scaling} and in particle-based numerical studies, in particular, in external non-linear potentials~\cite{dabelow2021irreversible} and interacting systems showing phase separation~\cite{crosato2019irreversibility, chiarantoni2020work}. Only, recently, we have derived analytical results for an interacting case, reporting the analytical expression for the entropy production of active solids formed by self-propelled particles~\cite{caprini2022entropons}, where simulations based on ABPs have been compared with theoretical results obtained through AOUPs.

The force between the particles is chosen as Eq.~\eqref{eq:linearforceapprox}, i.e. the system is assumed to be in solid-like configurations. In two dimensions, particles are arranged in a hexagonal lattice, as usual for systems of pure repulsive particles at high density, while a more complex scenario can occur in three dimensions.
To achieve active solid configurations, for instance with purely repulsive particles, one has to consider large packing fractions and small equilibrium fluctuations (small thermal temperature, for instance), but also small non-equilibrium fluctuations, controlled by the active temperature $T_a=v_0^2 \gamma \tau$.
Indeed, in two dimensions the increase of $T_a$ 
shifts the melting transition to larger densities~\cite{bialke2012crystallization, cugliandolo2018phases, digregorio2018full, caprini2020hidden, klamser2018thermodynamic, li2021melting, hopkins2022yield, omar2021phase}, inducing a fluidization of the system, and broadens the hexatic region~\cite{digregorio2018full, negro2022inertial, pasupalak2020hexatic}.
Active solids were explored mostly in one~\cite{caprini2020time, gupta2021heat, singh2021crossover, sandoval2022minimal, santra2022activity} and two-dimensions~\cite{hawkins2014stress, digregorio2018full, ophaus2021two, praetorius2018active, lin2021order} where they exhibit fascinating phenomena without a passive counterpart, displaying traveling crystals~\cite{menzel2013traveling, menzel2014active, briand2018spontaneously}, spatial velocity correlations~\cite{caprini2020hidden, henkes2020dense, caprini2021spatial}, collective rotations~\cite{ferrante2013elasticity, huang2020dynamical} as well as an intriguing scenario in the formation of topological defects~\cite{digregorio2022unified}.
However, before Ref.~\cite{caprini2022entropons}, collective excitations in active solids were poorly investigated and understood.

In Fourier space, the dynamics of crystal formed by self-propelled particles following the AOUP model reads
\begin{subequations}
\label{eq:fourierspace_nonpolarsolid}
\begin{align}
\label{eq:active_solid_eqmotion}
&\left(- m \omega^2 + i\omega\gamma +m \omega^2(\mathbf{q}) \right)\hat{\mathbf{u}}=\sqrt{2T\gamma} \hat{\boldsymbol{\xi}} + \gamma v_0 \hat{\mathbf{n}} \\
\label{eq:active_dynamicsn}
&\left( i \omega \tau+1\right) \hat{\mathbf{n}} = \sqrt{2 \tau}\hat{\boldsymbol{\eta}}
\end{align}
\end{subequations}
which compared with Eqs, \eqref{eq:dynamics_thermalsolid} contains an extra active force term $\hat{\mathbf{n}}= \hat{\mathbf{n}}(\omega, \mathbf{q})$.
To  identify phonons and entropons, we first recognize that $G(\omega, \mathbf{q})$, the response function, coincides with the equilibrium expression~\eqref{eq:Gthermal}.
As a consequence, the phonons of the active solids and those of the equilibrium crystal have the same correlation function given
by Eq.\eqref{eq:Ceq_thermal}.
{
The non-equilibrium force, $\gamma v_0 \mathbf{n}$, produces an additional contribution to the displacement correlation, 
the entropons, and
generates entropy production as shown by the relation
\begin{equation}
\label{eq:Cout_activesolid}
\frac{\mathcal{C}_{out}(\omega, \mathbf{q})}{T} =
\frac{\sigma(\omega, \mathbf{q})}{\omega^2\gamma} .
\end{equation}
Here, the spectral entropy production is given by
\begin{equation}
\label{eq:sigma_activesolid}
\sigma(\omega, \mathbf{q})=\frac{T_a}{T} \frac{K(\omega)}{\tau_I^2} \frac{\tau_I^2 \omega^2}{\tau_I^2(\omega^2-\omega^2(\mathbf{q}))^2 +\omega^2} \,.
\end{equation}
with $K(\omega)$ representing a Lorentzian shape function
\begin{equation}
K(\omega)=\frac{1}{1+\omega^2\tau^2}\,,
\end{equation}
with an explicit dependence on $\omega$ but not on  $\mathbf{q}$.
To calculate expression~\eqref{eq:sigma_activesolid}, we use Eq.~\eqref{eq:sigma_usefulformula}, which requires the knowledge of $G(\omega, \mathbf{q})$) as well as the dynamical correlations of the active force $\langle \mathbf{n}(\omega, \mathbf{q})\cdot\mathbf{n}(-\omega, -\mathbf{q})\rangle$ derived in Appendix~\ref{app:explicitcalculations}.

Entropons coexist with phonons~\cite{caprini2022entropons}, as independent collective excitations with strength proportional to the active temperature $T_a=v_0^2\gamma\tau$ and have the property of
vanishing at equilibrium when the active force also vanishes in the limits $v_0\to 0$ and/or $\tau\to0$.
By comparing the amplitudes of phonons and entropons, we realize that entropons play a negligible role, when the thermal temperature is larger than the active temperature, $T\gg T_a$, while entropons dominate over phonons in the opposite limit $T \ll T_a$.

$\mathcal{C}_{out}(\omega, \mathbf{q})$ as a function of frequency changes its shape according to the values of $\mathbf{q}$, the inertial time and the persistence time through $K(\omega)$.
At fixed $\tau_I \omega(\mathbf{q})$,  $K(\omega)$ kills the high frequency tails of $\mathcal{C}_{out}(\omega, \mathbf{q})$ when $\tau$ is large and shifts the peaks of $\mathcal{C}_{out}(\omega, \mathbf{q})$ for frequency smaller than the dispersion relation $\omega(\mathbf{q})$. We remark that for $\tau_I\omega(\mathbf{q}) \to \infty$ at fixed $\tau$, also $\mathcal{C}_{out}(\omega, \mathbf{q})$ becomes a $\delta$-function peaked at $\omega \sim \omega(\mathbf{q})$.

Finally, we comment that the theory based on the AOUP model, originally developed in Ref.~\cite{caprini2022entropons}, has been successfully compared with simulations of ABP solid-like phases, revealing a good agreement.

\subsection{Self-propelled solids with alignment interactions}

Several  active matter systems display collective behaviors: at the macroscopic scale, birds flock in the sky~\cite{cavagna2014bird}, fish display schooling~\cite{ward2008quorum}, while insects swarm in large clouds~\cite{cavagna2017dynamic}. Additionally, at the mesoscopic scale, cell monolayers~\cite{alert2020physical, malinverno2017endocytic} and bacteria~\cite{be2020phase} exhibit similar phenomena, and flocking behavior has been observed in active colloids, such as Quinke rollers \cite{bricard2013emergence, geyer2019freezing}.
These phenomena are usually reproduced through particle-based models involving the introduction of explicit forces responsible for the local alignment of the particles' orientations~\cite{vicsek2012collective}.
The first example of this approach dates back to the pioneering work of Vicsek~\cite{vicsek1995novel} and successively to variants of his model~\cite{chate2008modeling}, such as the inertial spin model~\cite{cavagna2015flocking} introduced to account for experiments showing the bird flocking.
Recently, the interplay between repulsive inter-particles forces and alignment interactions has been investigated~\cite{pu2017reentrant, D2SM00385F, knezevic2022collective} and shows a rich phenomenology displaying phase-separation, flocking clusters \cite{chate2008collective, martin2018collective, pu2017reentrant} and traveling bands~\cite{knezevic2022collective}.

Here, we include alignment between the orientations of the particles in the perhaps simplest way, i.e. through linear interactions between the orientational vectors of neighboring particles. This is a sensible assumption because particles are in a solid-like configuration.
Again, by using the AOUP dynamics  $\mathbf{n}_i$ evolves according to:
\begin{equation}
\tau\dot{\mathbf{n}}_i = - \mathbf{n} + \sqrt{2\tau} \boldsymbol{\eta}_i + \tau\alpha \sum^{*}_{j} (\mathbf{n}_j -\mathbf{n}_i)
\end{equation}
where $\alpha$ is a parameter determining the strength of the alignment and the sum, $\overset{*}{\sum}$, runs over first neighbors.
We have restricted the alignment interactions to nearest neighbors
but the method could easily include alignment interactions of next nearest neighbors.

Applying the double FT, the dynamics takes the form
\begin{subequations}
\begin{align}
&\left(- m \omega^2 + i\omega\gamma +m \omega^2(\mathbf{q}) \right)\hat{\mathbf{u}}=\sqrt{2T\gamma} \hat{\boldsymbol{\xi}} + \gamma v_0 \hat{\mathbf{n}} \\
\label{eq:polarsolid_orientation}
&\left( i \omega \tau+1 + \alpha \tau \omega^2(\mathbf{q})\right) \hat{\mathbf{n}} = \sqrt{2 \tau}\hat{\boldsymbol{\eta}}
\end{align}
\end{subequations}
which resembles Eq.~\eqref{eq:fourierspace_nonpolarsolid}, except for the mapping $1 \to 1 +\alpha \tau \omega^2(\mathbf{q})$ in the dynamics of $\hat{\mathbf{n}}(\omega, \mathbf{q})$, Eq.~\eqref{eq:polarsolid_orientation}.
As a consequence, we expect solutions formally similar to those obtained in the absence of alignment interactions, with a renormalization of the persistence time.
Upon identifying $G(\omega, \mathbf{q})$ with Eq.~\eqref{eq:Gthermal}, immediately we obtain the solution for $\mathcal{C}_{eq}(\omega, \mathbf{q})$ that coincides with that obtained for equilibrium and non-aligning active solids (Eq.~\eqref{eq:Ceq_thermal}).
As in the previous cases, the contribution of phonons is not affected by the active force and by the presence of alignment interactions.

Now, the expression of $\mathcal{C}_{out}(\omega, \mathbf{q})$ formally coincides with the one obtained for non-aligning active solids, i.e. Eq.~\eqref{eq:Cout_activesolid} with
the difference entirely contained in the spectral entropy production, which can be calculated by using Eq.~\eqref{eq:sigma_usefulformula} and, then, by estimating the dynamical correlation of the active force $\gamma^2v_0^2\langle \mathbf{n}(\omega, \mathbf{q}) \cdot \mathbf{n}(-\omega, -\mathbf{q})\rangle$ (see Appendix \ref{app:explicitcalculations}). In this way, we obtain
\begin{equation}
\label{eq:eq_entropyprod_activesolid_align}
\sigma(\omega, \mathbf{q})=\frac{T_a}{T} \frac{K_a(\omega, \mathbf{q})}{\left(1+ \tau\alpha\frac{\omega^2(\mathbf{q})}{\omega^2_E} \right)^2} \frac{ \omega^2}{\tau_I^2(\omega^2-\omega^2(\mathbf{q}))^2 +\omega^2} .
\end{equation}
As in the non-aligning active crystal, $\sigma(\omega, \mathbf{q})\propto T_a/T$ and contains the same term as the one featuring in Eq.~\eqref{eq:sigma_activesolid} and involving the difference $\omega^2 - \omega^2(\mathbf{q})$.
However, the presence of alignment interactions induce an additional dependence on the dispersion relation $\omega^2(\mathbf{q})$ and a renormalized shape function $K_a(\omega, \mathbf{q})$ given by
\begin{equation}
K_a(\omega, \mathbf{q})=\frac{\left(1+ \tau\alpha\frac{\omega^2(\mathbf{q})}{\omega^2_E} \right)^2}{\left(1+ \tau\alpha\frac{\omega^2(\mathbf{q})}{\omega^2_E} \right)^2+\omega^2\tau^2}\,.
\end{equation}
We remark that, in this case, also $K_a(\omega, \mathbf{q})$ depends explicitly on $\mathbf{q}$ through the dispersion relation $\omega^2(\mathbf{q})$.
Depending on the value of $\tau \alpha$, the shape function can significantly shift the typical frequency $\omega$ at which $K_a(\omega, \mathbf{q})$ assumes values smaller than 1. Its effect is conceptually similar to that of $K(\omega)$, since also $K_a(\omega, \mathbf{q})$ cuts the high frequencies as $\alpha$ increases.
As a consequence, the increase of $\alpha$ changes the position of the main peak of $\sigma(\omega, \mathbf{q})$ inducing a shift for smaller $\omega$ that significantly depends on $\mathbf{q}$, at variance with active solids without alignment interactions where the shift is $\mathbf{q}$-independent.






\subsection{Self-propelled solids with contact friction}

In systems of cell monolayers~\cite{alert2020physical, garcia2015physics, peglion2014adherens, czirok2013collective} as well as in granulars and active granulars~\cite{deseigne2010collective, scholz2018rotating}, 
particles exhibit contact friction forces that usually slow down the motion and give rise to local alignment of the particle velocity.
In this case, the particle dynamics reads
\begin{subequations}
\label{eq:contactfriction}
\begin{align}
&\dot{\mathbf{x}}_i=\mathbf{v}_i\\
&m\dot{\mathbf{v}}_i= -\gamma \mathbf{v}_i + \mathbf{F}^{lattice}_i +\sqrt{2 T \gamma} \boldsymbol{\xi}_i + \mathbf{f}^a_i + \mathbf{F}_c \,,
\end{align}
\end{subequations}
where the additional force $\mathbf{F}_c$ is given by
\begin{equation}
\mathbf{F}_c= m\gamma_c \sum_j^* (\mathbf{v}_j - \mathbf{v}_i) \,.
\end{equation}
Here, $\gamma_c$ represents the friction coefficient due to contact friction between particles, and the sum $\sum_j^*$ is restricted to the first neighbors of the particle $i$. This force induces the alignment between the particle velocities pushing $\mathbf{v}_i$ towards the average velocities of the neighboring particles.

The dynamics~\eqref{eq:contactfriction} can be easily expressed in Fourier space as
\begin{equation}
\begin{aligned}
&\left(- m \omega^2 + i\omega\gamma +m \omega^2(\mathbf{q}) \left[1+i\omega \gamma_c\right] \right)\hat{\mathbf{u}}\\
&=\sqrt{2T\gamma} \hat{\boldsymbol{\eta}} + \gamma v_0 \hat{\mathbf{n}}
\end{aligned}
\end{equation}
where $\hat{\mathbf{n}}$ evolves as Eq.~\eqref{eq:active_dynamicsn}.
After identifying the expression for $G(\omega, \mathbf{q})=\mathcal{R}_{uu}(\omega, \mathbf{q})$
\begin{equation}
G(\omega, \mathbf{q}) = \frac{1}{\left(- m \omega^2 + i\omega\gamma +m \omega^2(\mathbf{q}) \left[1+i\frac{\omega \gamma_c}{\omega^2_E}\right] \right)}
\end{equation}
and, consequently, $L(\omega, \mathbf{q})$ as its inverse, one determines the phonon contribution, i.e. the equilibrium part, to the dynamical correlation $\mathcal{C}_{eq}(\omega, \mathbf{q})$
\begin{equation}
\gamma\frac{\mathcal{C}_{eq}(\omega, \mathbf{q})}{T}= \frac{1}{\tau_I^2[\omega^2(\mathbf{q})-\omega^2]^2 + [1 + \gamma_c \tau_I \frac{\omega^2(\mathbf{q})}{\omega^2_E}]^2} \,.
\end{equation}
At variance with all cases above, contact friction interactions produce a shift in the spectrum of the thermally excited phonons. Such a shift depends on $\mathbf{q}$ since the velocity coupling term becomes larger as $\omega^2(\mathbf{q})$ is increased and has an amplitude determined by the contact friction coefficient $\gamma_c$.

We identify the contribution of entropons, due the presence of the active force $\mathbf{f}_i^a$, as 
the non-equilibrium part of the dynamical correlations of the displacement $\mathcal{C}_{out}(\omega, \mathbf{q})$:
\begin{equation}
\gamma\frac{\mathcal{C}_{out}(\omega, \mathbf{q})}{T} =\frac{\sigma(\omega, \mathbf{q})}{\omega^2 \left(1 + \gamma_c \tau_I \frac{\omega^2(\mathbf{q})}{\omega^2_E}\right)^2}
\end{equation}
where the spectral entropy production $\sigma(\omega, \mathbf{q})$ has the same expression as the one obtained in the case of active solids without alignment interactions, i.e. Eq.~\eqref{eq:sigma_activesolid}.
However, the contribution of entropons to the correlation function  is shifted by contact frictions: the system behaves as if was subject to an effective friction coefficient $\gamma+\gamma_c m \omega^2(\mathbf{q})/\omega^2_E$ that depends on the dispersion relation $\omega^2(\mathbf{q})$ and becomes larger as $\gamma_c$ increases.

\section{Conclusions}\label{sec:conclusions}

\subsection{Summary}

%

In this paper, we have generalized the concept of entropons originally introduced for active crystals formed by self-propelled particles in the absence of alignment interactions~\cite{caprini2022entropons}.
Here, we have shown that the picture of entropons is much more general and apply to a variety of out-of-equilibrium crystals where each particle of the solid is driven intrinsically.
This generality is demonstrated for a broad class of crystals reaching a non-equilibrium steady state and it is discussed explicitly for several examples, such as active solids formed by particles with alignment interactions or contact friction forces.
In these cases, the spectral entropy production, the dynamical correlations of the particle displacement, and its response function have been analytically calculated as a function of the model parameters. This corroborates the distinction between thermal phonons, excited by Brownian translational noise, and entropons originating from the intrinsic non-equilibrium nature of the dynamics and associated with the entropy production and, thus, violation of detailed balance.

\subsection{Discussion}

Entropons provide a link between solid-state physics and stochastic thermodynamics, showing how non-equilibrium observables such as entropy production are related to the formation of novel collective excitations.
The concept we are proposing is rather general: entropons characterize any non-equilibrium crystals and not only solids formed by self-propelled particles.
While we have explored in this paper the existence of entropons for diagonal systems, where different spatial components of the dynamics are not coupled in Fourier space, a theoretical challenge could be represented by the extension of our results to non-diagonal cases, where, for example, a magnetic field~\cite{vuijk2020lorentz, abdoli2020stationary} induces spontaneous rotations in the particle trajectories. Even more challenging is the case of non-equilibrium forces, which are odd under time-reversal~\cite{spinney2012nonequilibrium}, that in principle could lead to collective excitations with a different nature.

The fact that entropons occur in different non-equilibrium systems will facilitate their verification in future experiments, both at the micron and macroscopic scales.
At the micron scale, these experiments can in principle involve cell monolayers at high density~\cite{alert2020physical} that include contact friction forces.
Moreover, entropons are observable in two-dimensional crystals formed by active colloidal particles. Explicit examples are Janus particles, in the denser phase of a motility-induced phase-separated system~\cite{buttinoni2013dynamical, ginot2018aggregation, van2019interrupted} also known as ``living crystals''~\cite{palacci2013living, mognetti2013living}, or high-density suspensions of Quinke rollers~\cite{bricard2013emergence, geyer2019freezing}.
The anisotropic and alignment interactions studied in this paper are often relevant in these systems.
Another promising example involves complex plasma crystals~\cite{ivlev2012complex} which can be enriched by light-induced activity~\cite{nosenko2022two}.
Finally, at the macroscopic scale, entropons are observable in solids formed by active granular particles that have been recently experimentally realized by connecting neighboring particles by springs~\cite{baconnier2021selective}. Active granulars are often modeled by means of alignment torques~\cite{kumar2014flocking} which falls into the generalization reported in this paper.

Finally, we point out that entropons are qualitatively different from bosons peaks, for instance occurring in glasses or supercooled liquids. These soft modes originate from the absence of long-range translational order in the system, at variance with entropons that are predicted through an ideal theory based on elastic solids.
Understanding how entropons interfere with those boson peaks represents a promising future research line to shed light on novel aspects of non-equilibrium physics.

%


\appendix

\section{Definition of the Fourier transforms}\label{app:Fouriertransformdefinition}

In this appendix, we provide the definitions of the Fourier transforms of the dynamical variable of the system (displacement, velocity, non-equilibrium force, active force, and so on) in the domains of frequency $\omega$ and wave vector $\mathbf{q}$.
For the sake of notational convenience, we denote the Fourier transform of a variable by a tilde. They are obtained by applying the operator
\begin{equation}
\lim_{t_w\to\infty}\int_{-t_w/2}^{t_w/2} dt\sum_{i=1}^{N} e^{- i \mathbf{q}\cdot \mathbf{x}^0_i} e^{- i \omega t}
\end{equation}
to a dynamical variable.
In particular, the Fourier transform of the particle displacement with respect to its unperturbed position in the lattice, i.e. $\mathbf{u}_i=\mathbf{x}_i-\mathbf{x}_i^0$, and that of the general non-equilibrium force $\mathbf{F}_i(t)$ are given by
\begin{subequations}
\begin{align}
\label{eq:displacement_fourier}
&\hat{\mathbf{u}}(\omega, {\mathbf{q}})=\lim_{t_w \to \infty}\int_{-t_w/2}^{t_w/2} dt\sum_{i=1}^{N} \mathbf{u}_i e^{- i \mathbf{q}\cdot \mathbf{x}^0_i} e^{- i \omega t} \\
&\hat{\mathbf{F}}(\omega, {\mathbf{q}})=\lim_{t_w \to \infty}\int_{-t_w/2}^{t_w/2} dt\sum_{i=1}^{N} \mathbf{F}_i e^{- i \mathbf{q}\cdot \mathbf{x}^0_i} e^{- i \omega t} \,. 
\end{align}
\end{subequations}
where the time used to define the Fourier transform, $t_w$, in practice, corresponds to the time window of the simulations.

\section{Expressions for the Einstein frequency of the solid}\label{app:dispersionrelation}

The dispersion relation of a solid $\omega(\mathbf{q})$ and the Einstein frequency $\omega_E$ depend on the dimension of the system and on the type of lattice where the particles organize.
To fix ideas, here, we report the expressions for $\omega(\mathbf{q})$ in several cases of interest, defining $\bar{r}$ as the lattice constant, i.e. the average distance between neighboring particles.
\begin{itemize}

\item For a $d$-dimensional solid, characterized by a square/cubic lattice the dispersion relation is given by
\begin{equation}
\omega^2(\mathbf{q})=2d\omega_E^2 \left(1- \cos{\left(q\bar{r}\right)} \right)
\end{equation}
while the Einstein frequency reads
\begin{equation}
\omega_E^2 =\frac{1}{2m} U''(\bar{r}) \,.
\end{equation}
Here, each prime denotes a derivative with respect to the argument of the potential calculated at $\bar{r}$.

\item
In the two-dimensional case,  where particles typically arrange on a triangular lattice, the dispersion relation is
\begin{equation}
\omega^2(\mathbf{q})=2 \omega_E^2 \left[ 3 - \cos{\left(q_x \bar{r}\right)} 
-2  \cos{\left(\frac{q_x}{2} \bar{r}\right)} 
\cos{\left(\frac{\sqrt{3}}{2}q_y \bar{r}\right)} 
\right]
\end{equation}
while the Einstein frequency of the solid reads
\begin{equation}
\omega_E^2 = \frac{1}{2m} \left( U''(\bar{r}) + \frac{U'(\bar{r})}{\bar{r}}\right) \,.
\end{equation}


\end{itemize}

\section{Explicit calculation of the entropy production for active solids with and without alignment interactions}\label{app:explicitcalculations}

In this appendix, we report the explicit calculation for spectral entropy production $\sigma(\omega, \mathbf{q})$ in the case of self-propelled (active) solids with and without alignment interactions between the active forces.
As shown in the Sec.\ref{sec:entropons} (Eq.~\eqref{eq:explicit_sigma_expression}), $\sigma(\omega, \mathbf{q})$ can be expressed as
\begin{equation}
\sigma(\omega, \mathbf{q})= \lim_{t\to\infty}\frac{i}{t} \frac{ \text{Im}[L(\omega, \mathbf{q})] }{2 T \gamma}\langle \hat{\mathbf{u}}(\omega, {\mathbf{q}})\hat{\mathbf{F}}(-\omega, -{\mathbf{q}})\rangle + c.c
\end{equation}
for the class of models that we have studied in this paper.
In the case of active solids, since $L(\omega, \mathbf{q})$ is given by
\begin{equation}
L(\omega, \mathbf{q}) = m \omega^2(\mathbf{q})- m \omega^2 + i \omega \gamma
\end{equation}
we have
\begin{equation}
\text{Im}[L(\omega, \mathbf{q})]=\omega \gamma \,.
\end{equation}
The Fourier transform of the general force $\mathbf{F}(\omega, \mathbf{q})$ can be identified with the Fourier transform of the active force $\gamma v_0 \hat{\mathbf{n}}(\omega, \mathbf{q})$.

By using the equation of motion for $\hat{\mathbf{u}}(\omega, \mathbf{q})$, i.e. Eq.~\eqref{eq:active_solid_eqmotion}, and that $\langle \boldsymbol{\xi}(\omega,\mathbf{q})\cdot \mathbf{F}(\omega, \mathbf{q})\rangle=0$, we obtain
\begin{equation}
\label{eq:app_sigma_active}
\sigma(\omega, \mathbf{q})= \lim_{t\to\infty} \frac{1}{t}\frac{\gamma^2 v_0^2}{2 T}\frac{i \omega\langle \hat{\mathbf{n}}(\omega, {\mathbf{q}})\hat{\mathbf{n}}(-\omega, -{\mathbf{q}})\rangle}{m \omega^2(\mathbf{q})- m \omega^2 + i \omega \gamma} + c.c \,.
\end{equation}
The dynamical correlation $\langle \hat{\mathbf{n}}(\omega, {\mathbf{q}})\hat{\mathbf{n}}(-\omega, -{\mathbf{q}})\rangle$ is calculated by using the dynamics of the active force in the AOUP model ( Eq.~\eqref{eq:polarsolid_orientation}) to obtain
\begin{equation}
\hat{\mathbf{n}}(\omega, \mathbf{q}) = \frac{\sqrt{2 \tau}\hat{\boldsymbol{\eta}}(\omega, \mathbf{q})}{\left( i \omega \tau+1 + \alpha \tau \omega^2(\mathbf{q})\right)} \,.
\end{equation}
and by multiplying the result by $\hat{\mathbf{n}}(-\omega, -\mathbf{q})$ and taking the average over the noise:
\begin{equation}
\begin{aligned}
&\langle \hat{\mathbf{n}}(\omega, \mathbf{q})\hat{\mathbf{n}}(-\omega, -\mathbf{q})\rangle = \frac{2 \tau \langle\hat{\boldsymbol{\eta}}(\omega, \mathbf{q})\hat{\boldsymbol{\eta}}(-\omega, -\mathbf{q})\rangle}{\left( \omega^2 \tau^2+(1 + \alpha \tau \frac{\omega^2(\mathbf{q}))^2}{\omega_E^2}\right)} \,,\\
& =2 \tau \langle\hat{\boldsymbol{\eta}}(\omega, \mathbf{q})\hat{\boldsymbol{\eta}}(-\omega, -\mathbf{q})\rangle
\frac{K_a(\omega, \mathbf{q})}{\left(1+ \tau\alpha\frac{\omega^2(\mathbf{q})}{\omega^2_E} \right)^2}
\end{aligned}
\end{equation}
Using the property $\langle \boldsymbol{\eta}(\omega, \mathbf{q})\cdot \boldsymbol{\eta}(\omega', \mathbf{q}') \rangle = \delta(\omega+\omega')\delta(\mathbf{q}+\mathbf{q}')$ that cancels out the term $\lim_{t\to \infty} 1/t$, we get
\begin{equation}
\begin{aligned}
\sigma(\omega, \mathbf{q})&= \frac{T_a}{T}\frac{K_a(\omega, \mathbf{q})}{\left(1+ \tau\alpha\frac{\omega^2(\mathbf{q})}{\omega^2_E} \right)^2}\frac{i \omega \gamma}{m \omega^2(\mathbf{q})- m \omega^2 + i \omega \gamma} + c.c\\
&= \frac{T_a}{T}\frac{K_a(\omega, \mathbf{q})}{\left(1+ \tau\alpha\frac{\omega^2(\mathbf{q})}{\omega^2_E} \right)^2} \frac{\omega^2}{\tau_I^2 (\omega_q^2-\omega^2)^2+\omega^2}
\end{aligned}
\end{equation}
that concides with the final expression for $\sigma(\omega,\mathbf{q})$, i.e. Eq.~\eqref{eq:sigma_activesolid} for $\alpha=0$ (such that $K_a \to K$) or Eq.~\eqref{eq:eq_entropyprod_activesolid_align} with $\alpha\neq0$.

\begin{acknowledgments}
\textit{Acknowledgments --- } 
LC acknowledges support from the Alexander Von Humboldt foundation.
UMBM  
acknowledges support from the MIUR PRIN 2017 project 201798CZLJ. 
HL acknowledge support by the Deutsche Forschungsgemeinschaft (DFG) through the SPP 2265 under the grant number LO 418/25-1.
\end{acknowledgments}

\bibliographystyle{apsrev4-1}

\bibliography{EP}

\end{document}